\documentclass[letter]{aa}
\usepackage{amsmath,bm}
\usepackage{amsfonts}
\usepackage{amssymb}
\usepackage{graphicx}
\usepackage{natbib}
\usepackage{scalerel}
\usepackage{csquotes}
\usepackage{lscape}
\usepackage{siunitx}
\usepackage{lastpage}

\usepackage[table]{xcolor}

\usepackage{txfonts}
\usepackage[pdfencoding=auto,psdextra]{hyperref}
\hypersetup{
    colorlinks=true,
    linkcolor=blue,
    filecolor=magenta,      
    urlcolor=blue,
    citecolor=blue
}
\makeatletter
\renewcommand*\aa@pageof{, page \thepage{} of \pageref*{LastPage}}
\makeatother

\usepackage[utf8]{inputenc}

\usepackage[switch, modulo]{lineno}
\linenumbers

\newcommand{\Planck}{\textit{Planck}}
\newcommand{\Herschel}{\textit{Herschel}}
\newcommand{\Spitzer}{\textit{Spitzer}}
\newcommand{\Euclid}{\textit{Euclid}}

\newcommand{\FT}[1]{
	{\rm FT} \left[ #1 \right]
}

\newcommand{\IFT}[1]{
	{\rm FT}^{-1} \left[ #1 \right]
}

\newcommand{\vect}[1]{\ensuremath{\bm{\hat{#1}}}}
\newcommand{\mean}[2][]{\ensuremath{\overline{#2}_{#1}}}
\newcommand{\meanKnull}{\mean[1012]{\kappa}}
\newcommand{\meanKPC}{\mean[\rm pc]{\kappa}}

\begin{document}

\title{Assessing the total mass of the \Planck\ protocluster candidates with CMB lensing stacking}
\author{
T.~Dusserre\inst{\ref{aff1},\ref{aff3}}\thanks{\email{tanguy.dusserre@universite-paris-saclay.fr}}
\and
H.~Dole\inst{\ref{aff1},\ref{aff5}}
\and
M.~Langer\inst{\ref{aff2}}
\and
G.~Fabbian\inst{\ref{aff1}}
}
\institute{Université Paris-Saclay, CNRS, Institut d'Astrophysique Spatiale, 91405 Orsay, France\label{aff1}
\and
LUX, Observatoire de Paris, Université PSL, Sorbonne Université, CNRS, 75014 Paris, France\label{aff2}
\and
Centre National d’Etudes Spatiales (CNES), 2, Place Maurice Quentin, 75039, Paris, France\label{aff3}
\and
Institut universitaire de France (IUF), 1 rue Descartes, 75231 PARIS CEDEX 05, France\label{aff5}
}
\newcommand{\orcid}[1]{}
\abstract{
Galaxy clusters in formation, or protoclusters, are important sites regarding the evolution of large scale structures and the star formation history of the Universe. With a few exceptions, only their member galaxies have been routinely observed so far. The majority of their mass is expected to be under the form of non baryonic dark matter and cold intergalactic gas.
}{
The \Planck\ high redshift star-forming field sample compiles 2151 protocluster candidates, among which 1012 fall within the footprint of the Atacama Cosmology Telescope (ACT). The lensing of the Cosmic Microwave Background (CMB) by foreground structures allows us to try and assess the total mass of these protoclusters for the first time.
}{
We stack the CMB lensing data from ACT at the location of the \Planck\ protocluster candidates to obtain a mean convergence signal. We then provide an estimate for the mean mass along the line of sight for these objects.
}{
The centre of the stacked field corresponds to a lensing signal at a $12\,\sigma$ significance level. The integration of the mass along the line of sight yields a value of $(6.6\pm 1.4)\times 10^{14}\, M_\odot$, similar to what is expected from numerical simulations. With the constraint $1.5<z<3$ from \Planck, we are able to rule out the possibility that this sample is populated by virialised massive structures.
}{ 
This is the first time that the mass of \Planck\ high redshift sources is measured in such a direct way. The results we obtain constitute a new strong argument in favour of this sample being composed of  a majority of galaxy protoclusters lying at $z \geq 1$. 
}

\keywords{Methods: statistical -- Surveys -- Cosmology: observations -- large-scale structure of Universe -- Cosmology: cosmic background radiation -- Galaxies: clusters: general}

\titlerunning{Assessing the total mass of the \Planck\ protocluster candidates with CMB lensing stacking}
\authorrunning{T. Dusserre et al.}
\maketitle
\nolinenumbers

\section{\label{sect:intro}Introduction}

Galaxy clusters are the largest gravitationally bound structures in the Universe, with sizes of the order of 1 Mpc and masses over $10^{14} M_\odot$. They originate from the overdensities seeded by quantum fluctuations at the end of inflation, which eventually collapse into massive haloes billions of years later. Such objects are expected to form at the nodes of the Cosmic Web \citep{Bond1996} by accreting matter through the filaments. 

However, there must have been a point in the history of the Universe where the components of future galaxy clusters were falling towards the centre of potential wells without having reached equilibrium yet. Such objects are called protoclusters and are usually defined simply as gravitationally bound dark matter, gas and galaxies that will virialise into a structure of mass superior to $10^{14}\, M_\odot$ at $z=0$ \citep[see][for a summary]{Overzier2016,alberts22}. Protoclusters were originally studied as groupings of galaxies around quasars, then many of them were also discovered in pointing observations \citep[see for instance ][]{koyamaPlanckselectedDustyProtocluster2021,laporteLensedProtoclusterCandidate2022,mascoloFormingIntraclusterGas2023,liEPOCHSPaperEnvironmental2024}. They usually lie at $z>1.5$ and some of their galaxies show clear signs of starburst activity \citep{flores-cacho2016,polletta2021,polletta2022}, or are expected to show such signs \citep{chiang2017}, hence the hypothesis that they play a key role in the star formation peak of the Universe, the Cosmic Noon \citep[see for instance][]{madau_cosmic_2014}. The protocluster redshift range is quite large, as protoclusters have been detected up to $z\simeq 8$ \citep{brinchCOSMOS2020IdentificationHighz2023,hashimoto2023}. So far, the proto-intracluster medium has only been detected in a few exceptional protoclusters \citep{daddi2022ApJ,mascoloFormingIntraclusterGas2023,zhou2026,bogdan2026}, the vast majority of them being characterised only by their member galaxies. 

Using these characteristics as a guide to find protoclusters, a catalogue of 2151 high redshift star forming sources was built \citep{planck16} with data from the \Planck\ all-sky survey \citep{planck2020_I}. Follow-up submillimetre and infrared observations with \Herschel\ and \Spitzer\ \citep{planck15,martinache2018} revealed that these fields of view concentrate multiple active galaxy overdensities, while simulations determined that about 72\% of these structures can be expected to evolve into galaxy clusters at $z=0$ \citep{gouinQuestioningPlanckSelected2022}. New cosmological surveys in the optical and near-infrared such as \Euclid\ \citep{euclid_aussel_2025} are the next step towards a full in-depth analysis of the galaxies of these protoclusters \citep[][]{dusserre2026}. However, as protoclusters are expected to be mainly composed of dark matter and cold gas \cite[e.g.][]{dekel06,daddi2022}, other strategies have to be deployed in order to estimate their total mass. 

\begin{figure*}[htbp!]
	\centering
	\includegraphics[width=0.85\hsize,trim={2cm 1cm 0.5cm 1cm},clip]{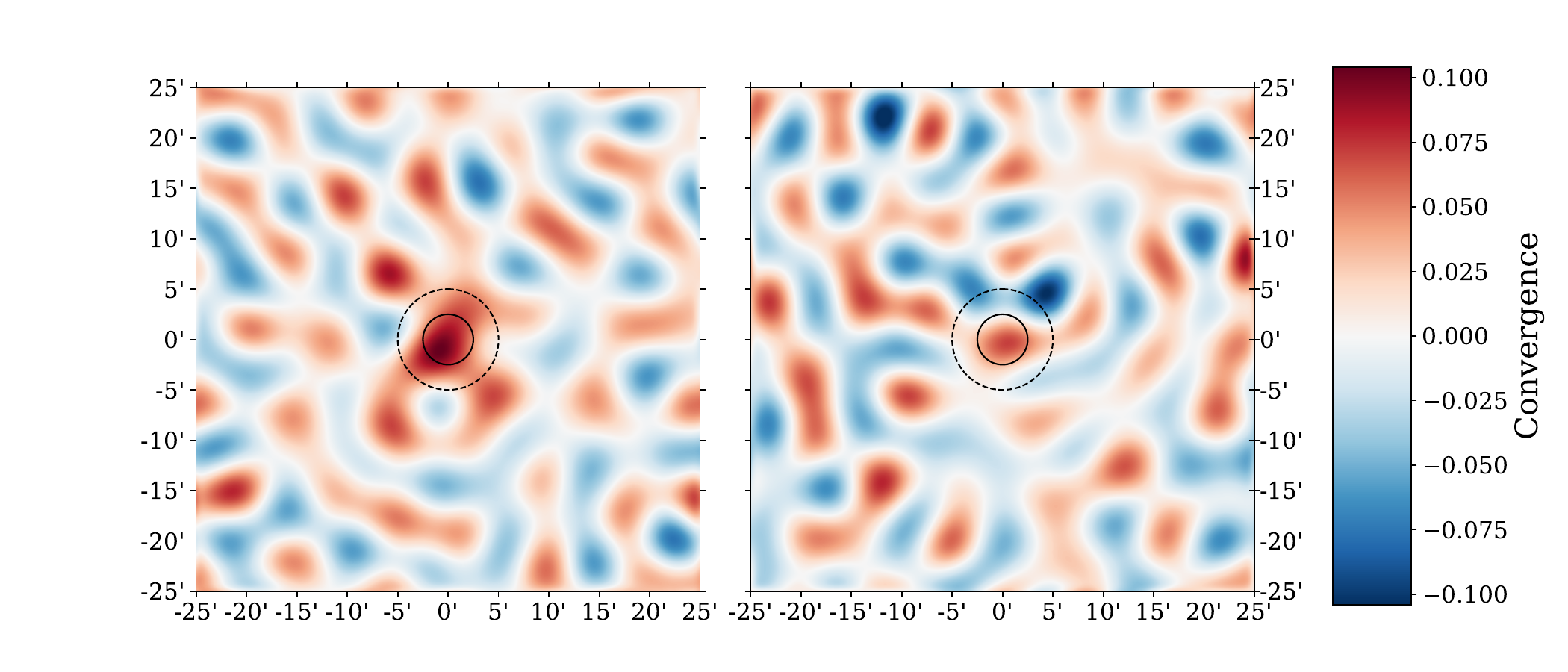}
	\caption{
    \ang{;50;}$\times$\ang{;50;} stacked ACT convergence maps smoothed by a Gaussian filter with a \ang{;3.6;} full width at half maximum for a better visualisation. On the left, we plot the stacked convergence map of the 1012 \Planck\ protocluster candidates covered by ACT.  A clear maximum in convergence and thus in surface mass density is visible at the location of the sources identified by \Planck. On the right, we show an example of null stacked convergence map over 1012 random positions in ACT's footprint. The signal in the centre circle of \ang{;5;} diameter, slightly larger than the \Planck\ beam, is a characteristic fluctuation due to noise. The noise is measured outside the dashed circle of \ang{;10;} diameter.\\
	\label{fig:convergence_map}
    }
\end{figure*}

Weak gravitational lensing astutely exploits the linear regime of general relativity to obtain quantities characterising the mass distribution on the line of sight. In particular, under the thin lens approximation, the convergence $\kappa(\bm{\hat{n}})$ for an observer at $z_{\rm obs}=0$ is directly related to the surface density $\Sigma (\bm{\hat{n}})$ in the direction $\bm{\hat{n}}$ (Eq.~\ref{eq:convergence}), normalised by a critical surface density $\Sigma_c(z_l,z_s)$ (Eq.~\ref{eq:critical_density})
\begin{subequations}
    \begin{eqnarray}
        \label{eq:convergence}
	    \kappa (\bm{\hat{n}})&=& \frac{\Sigma(\bm{\hat{n}})}{\Sigma_c(z_{\rm obs},z_l,z_s)}\\
        \label{eq:critical_density}
        \Sigma_c(z_{\rm obs},z_l,z_s) &=& \frac{c^2}{4\pi G_{\rm N}} \frac{D_A(z_{\rm obs},z_s)}{D_A(z_{\rm obs},z_l) D_A(z_l,z_s)} \, \cdot
    \end{eqnarray}
\end{subequations}

Here $G_{\rm N}$ is the Newtonian gravitational constant, $c$ is the speed of light, $D_A$ is the angular diameter distance between two redshifts, $z_l$ is the redshift of the lens, and $z_s$ is the redshift of the source. The last three quantities depend on a cosmological model, so $\Sigma_c(z_l,z_s)$ is sensitive to cosmological parameters. In the case of CMB lensing, the source is at $z_s = z_{\rm CMB} = 1090$. The evolution of $\Sigma_c(z_l,z_{\rm CMB})$ with respect to $z_l$, assuming the cosmology from \cite{planck2020_I} and \citet{planck2020_VI}, is plotted on Fig.~\ref{fig:crit_density}. For a detailed derivation, the reader can refer to \cite{bartelmannWeakGravitationalLensing2016}. In our case, the CMB acts as a background source even for protoclusters. CMB lensing maps have been reconstructed from the \Planck\ data in the past \citep{planck2014_XVIII} and more recently from the Atacama Cosmology Telescope (ACT) data \citep{ACT2024}, showing in both cases a good correlation with the Cosmic Infrared Background. CMB lensing has already been confirmed as a promising proxy to measure the mass of high redshift structures \citep{Madhavacheril2020,Huchet2024}, including galaxy protoclusters \citep{Gardner2024}.

In this letter, we present a complementary study of the \Planck\ protocluster candidates from \citet{planck16} by constraining their total mass through CMB lensing. Throughout this work, we use the flat $\Lambda$CDM cosmology from \citet{planck2020_I} and \citet{planck2020_VI}, with $H_0$ = 67.66 km.s$^{-1}$.Mpc$^{-1}$, $\Omega_{{\rm m},0}$ = 0.3111, $\Omega_{\Lambda,0}=0.6889$, $\Omega_{{\rm b},0}=0.4897$ $\sigma_8=0.8102$, and $n_s=0.9665$.

\section{\label{sect:data}ACT CMB lensing data}

Products of the observations made by ACT and processed through its pipeline are publicly available\footnote{\url{https://lambda.gsfc.nasa.gov/product/act/actadv_prod_table.html}}. For this study, we use the convergence maps in the sixth data release, described in \citet{qu2024,ACT2024}. ACT observations are ground-based and the footprint covers 23\% of the sky. \Planck\ measured the CMB anisotropies on the whole sky from the L2 point in space. Of the 2151 protocluster candidates detected on 28\% of the sky, 1012 lie in the part of the sky observed by ACT.

The ACT lensing data does not consist in the convergence map, but in its spherical harmonics. These were computed for multipoles $40\leq \ell \leq 3000$, which is equivalent to an angular resolution of about \ang{;3.6;}. Our knowledge of the angular size of the \Planck\ protocluster candidates is limited by the transfer function of \Planck, so we consider their angular diameter to be at most \ang{;5;}, which is the full width at half maximum of this telescope's beam. Therefore, the better angular resolution of ACT CMB lensing maps should be high enough to study the mass distribution of \Planck\ protocluster candidates. 

\section{\label{sect:methods}Stacked convergence map of the \Planck\ protocluster candidates}

Several convergence maps reconstructed from ACT data are publicly available. To obtain a high enough signal to noise ratio, we use the maps reconstructed from both temperature and polarisation anisotropies. However, temperature maps are sensitive to foreground emissions, and \Planck\ protocluster candidates are strong emitters in the far infrared. Therefore, we choose to use the ACT Cosmic Infrared Background (CIB) deprojected maps to avoid potential biases in the convergence estimate. 

We use the python package \texttt{healpy}\footnote{\url{https://healpy.readthedocs.io/en/latest/}} to compute the convergence map from the spherical harmonics and to obtain its tangent projection around every protocluster candidate. The flat coordinate grid of each \ang{;50;}$\times$\ang{;50;} zone centred around a \Planck\ detection is related to the \texttt{HEALPix} pixel mapping of the ACT convergence map (using \texttt{ang2pix}). At this stage, we have 1012 maps, each \ang{;50;}$\times$\ang{;50;} centred on a protocluster candidate.
We then take the average of all these maps to obtain a stacked, effective \Planck\ protocluster candidate, on which we do our analysis. A secondary analysis in which we take the median rather than the average yields practically identical results. The zone of the resulting image outside the centred disk of diameter $\ang{;10;}$ is used for the computation of noise statistics (Fig.~\ref{fig:convergence_map}). Lensing estimators generally have a response to null signal caused by boundaries and masks, which must be subtracted from the estimated map \citep[see for instance][]{planck2018_VIII}. We compute this zero-point convergence using the simulations provided by \cite{farren2024}.

The transfer function of ACT, the map reconstruction from spherical harmonics and its projection on the sky plane induce correlations between nearby pixels. We estimate the covariance directly from the stacked map by simulating 2000 maps with the same statistics as the background noise (see Appendix \ref{appendix:noise} for details). By doing so, we have 2000 different random estimations for each pixel, from which we compute the covariance of the map. This allows us to estimate the variance due to the noise inside the circle of diameter \ang{;5;} at the centre. 

\section{Mass signal at the location of \Planck\ protocluster candidates}

\subsection{\label{sect:results_convergence}Stacked ACT  convergence maps}

We present in Fig.~\ref{fig:convergence_map} the stacked convergence map of the 1012 \Planck\ protocluster candidates in ACT's footprint. The zone in which those candidates lie is delimited by a solid circle at the centre of diameter \ang{;5;}  Inside, convergence reaches a global maximum of 0.124 and has a mean value of $\meanKPC=0.082$. For comparison, the noise computed outside  the circle of diameter \ang{;10;} has a mean value 0.002 and a standard deviation of 0.031.

\begin{figure}
    \centering
    \includegraphics[width=0.9\hsize]{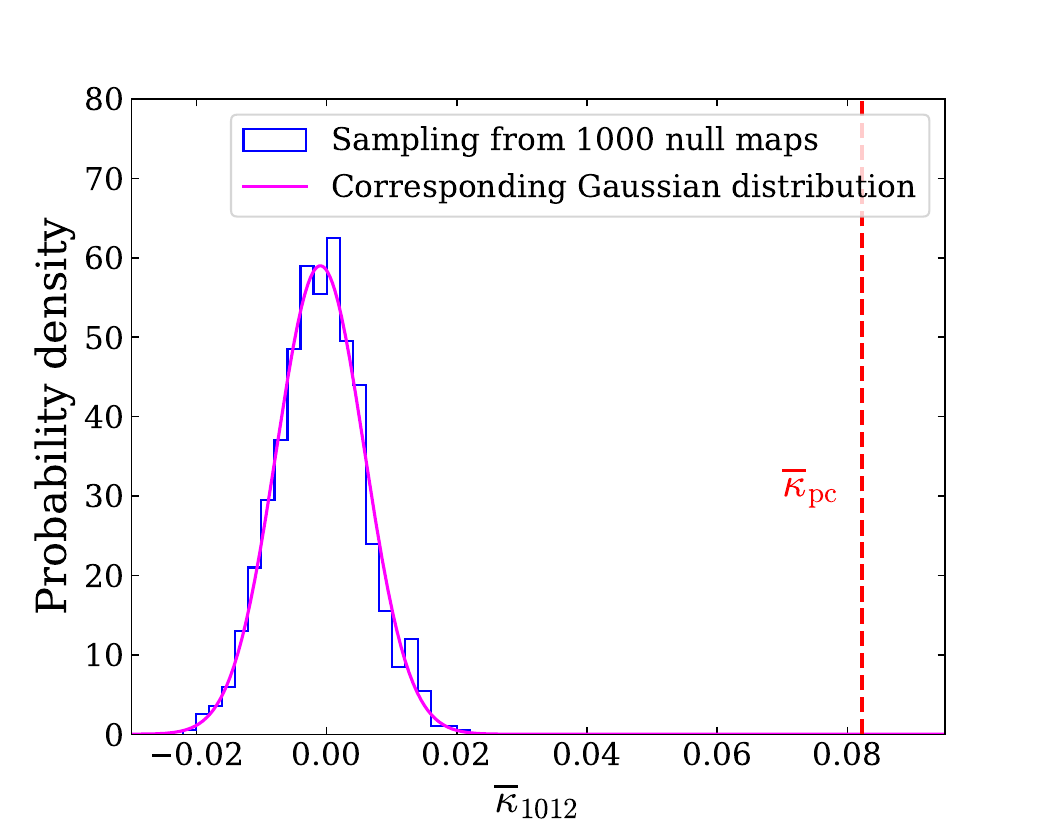}
    \caption{We display in blue the sampled probability distribution for the mean convergence value \meanKnull\ of a stack of 1012 random fields in ACT footprint. The gaussian of corresponding mean and standard deviation is plotted in magenta and is a satisfying representation of the blue histogram. The dashed red vertical line is the mean convergence value of our signal \meanKPC\ and corresponds to a $12\,\sigma_{\meanKnull}$ significance level.}
    \label{fig:mean_kappa}
\end{figure}

To quantify the significance level of our signal, we realise 1000 null maps by stacking each time 1012 random fields in ACT's footprint. We display a realisation of one of these null maps on the right of Fig.~\ref{fig:convergence_map}. This allows us to sample the probability distribution for the mean convergence value \meanKnull\ of a stack of 1012 random fields inside a circle of angular diameter \ang{;5;}. The sampled probability distribution we obtain (Fig.~\ref{fig:mean_kappa}) is pretty close to a Gaussian one, which can be expected as the null maps should be practically uncorrelated given the typical angular distance between two \ang{;5;} diameter circles. It has a mean value of $\langle \meanKnull\rangle=-0.9\times 10^{-3}$ and a standard deviation of $\sigma_{\meanKnull}=6.8\times 10^{-3}$. The mean convergence \meanKPC\ of the effective \Planck\ protocluster candidate therefore corresponds to a significance level of $\meanKPC-\langle \meanKnull\rangle=12\,\sigma_{\meanKnull}$, making it clearly different from the lensing reconstruction and sampling noises. 

\subsection{\label{sect:results_mass}Estimating the projected mass}

The convergence is related to the mass surface density by Eq.~\ref{eq:convergence} (cf Sect.~\ref{sect:intro}). In order to assign a mass $M_{\rm pc}(z)$ to the stacked, effective \Planck\ protocluster candidate, we simply formulate the hypothesis that all 1012 sources lie at the same redshift $z$. Under this hypothesis, $M_{\rm pc}(z)$ is given by Eq.~\ref{eq:mass} :

\begin{equation}
	\label{eq:mass}
	M_{\rm pc}(z) = \Sigma_c(z_{\rm obs},z,z_{\rm CMB}) D^2_A(z_{\rm obs},z) \int_{\Omega}  \kappa (\vect{n}) {\rm d}^2 \vect{n} \, \cdot
\end{equation}

To take into account the uncertainty on $z$, we consider a uniform distribution over the range $[1.5, 3]$, in which the \Planck\ protocluster candidates are expected to lie \citep{planck16}. Under these conditions, the mean value for the mass is $6.61\times 10^{14}\, M_\odot$, and the standard deviation is $0.26\times 10^{14}\, M_\odot$. Therefore, the assumption of a particular value for the redshift only affects the mass estimation by 4\%. Using the method described in Sect.~\ref{sect:methods}, we derive a $1.35\times 10^{14}\,M_\odot$ uncertainty due to the noise. Combining the uncertainties from noise and redshift, the mass estimate of the effective \Planck\ protocluster candidate is $M_{\rm pc} = (6.61\pm 1.37)\times 10^{14}\, M_\odot$. Estimates made with lensing maps reconstructed from temperature only, polarisation only, and temperature and polarisation yield results consistent with this one at $1\,\sigma$ (Appendix~\ref{appendix:other_maps}).

\section{\label{sect:discussion_protocluster}Concerning the nature of \Planck\ protocluster candidates}

Since the mean convergence value at the location of \Planck\ protocluster candidates is very unlikely to emerge from noise fluctuations (Sect.~\ref{sect:results_convergence}), we can safely assume that it is a true lensing signal caused by massive objects along the line of sight. \Planck\ estimations for the redshift of these structures rule out the possibility that they are situated at $z < 1$ \citep{planck16}.

The most probable astrophysical objects with a mass comparable to that computed in Sect.~\ref{sect:results_mass} are galaxy clusters and protoclusters. The fact that they lie at $z>1$ together with their mass estimate is in favour of the second possibility. We can quantify the probability $p(M,z)$ that an object of mass at least $M$ at redshift $z$ is a collapsed and virialised halo using the Press-Schechter formalism \citep[][see Appendix \ref{appendix:press_schechter} for details]{press1974}. The probability that no object of mass $M$ is a halo among a sample composed of $N=1012$ objects of such mass is then $\mathbb{P}_N(M,z)=[1-p(M,z)]^N$. Using \texttt{CAMB}\footnote{\url{https://camb.readthedocs.io/en/latest/}} to get $\sigma(M,z)$, we compute these quantities for $z=\{1.5,2,3\}$ (Tab.~\ref{tab:halo_probas}). We set $M=5 \times 10^{14}\,M_\odot$, which is the lower limit for the mass of the effective \Planck\ protocluster we obtained. All probabilities to form such massive haloes at the considered redshifts are extremely low, as the collapsing threshold requires at least a $4.8 \, \sigma(M,z)$ fluctuation of the linear overdensity field. The probability that even one single object is virialised is lower than $2 \times 10^{-3}$. This strongly disfavours the possibility that these objects are galaxy clusters at equilibrium.

\renewcommand{\arraystretch}{1.2}

\begin{table}[htbp!]
    \centering
    \caption{Probability that virialised haloes of mass at least $M$ have formed at the same redshift as \Planck\ protocluster candidates.}
    \begin{tabular}{ccccc}
    \hline\hline
    $z$ & $\sigma(M,z)$ & $\nu(M,z)$ & $p(M,z)$ & $\mathbb{P}_N(M,z)$ \\
    \hline
    $1.5$ & $0.351$ & $4.80$ & $1.61 \times 10^{-6}$ & $1-1.6 \times 10^{-3}$ \\
    $2$ & $0.296$ & $5.70$ & $1.22\times 10^{-8}$ & $1-1.2 \times 10^{-5}$\\
    $3$ & $0.224$ & $7.53$ & $5.14 \times 10^{-14}$ & $1-5.2 \times 10^{-11}$ \\
    \hline\hline
    \end{tabular}
    \tablefoot{For different redshifts $z$, we display the standard deviation of the smoothed matter overdensity field $\sigma(M,z)$, the peak height parameter $\nu(M,z)=\delta_c/\sigma(M,z)$, the probability $p(M,z)$ that a halo of mass at least $M$ has collapsed, and the probability $\mathbb{P}_N(M,z)$ that there are no haloes among 1012 regions of mass $M$. We take $M=5 \times 10^{14}\,M_\odot$ here.}
    \label{tab:halo_probas}
\end{table}

\citet{gouinQuestioningPlanckSelected2022} showed that 72\% of the most star-forming haloes between $z=1.5$ and $z=3$ in the TNG300 cosmological simulation end up inside galaxy clusters of masses superior to $10^{14}\,M_\odot$ at $z=0$. The \Planck\ protocluster candidate PHz G237.01$+$42.50 at $z=2.16$ is expected to evolve into a galaxy cluster of $6\times10^{14}\,M_\odot$ \citep{polletta2021}. Millimetre spectroscopic observations of PHz G037.4$-$57.4 \citep{kneissl2019,hill2024} and optical and near infrared photometric observations of twenty counterparts to the eight \Planck\ candidates lying in the Euclid Q1 field \citep{dusserre2026} also point towards present day masses of at least $10^{14}\,M_\odot$ in the redshift range $[1.5,3]$ for the majority of the sample. These values are compatible with our estimation of the mean mass in Sect.~\ref{sect:results_mass}.

 If these objects are single protoclusters, they will become among the largest clusters at $z=0$. In this case, we would expect them to be situated at the confluence of several cosmic filaments, which might eventually be detected with Ly\,$\alpha$ or 21\,cm observations. Another possibility is that they consist in alignments of several protoclusters with total masses of a few $10^{14} M_\odot$ along the line of sight. It has already been demonstrated that at least some of these fields contain multiple overdensities with star-forming galaxies \citep[][]{planck15,flores-cacho2016,negrello2017,gouinQuestioningPlanckSelected2022,dusserre2026}. Finally, as this mass was derived by integrating the mean surface density, this result might indicate the presence of a few very massive structures in the \Planck\ protocluster candidates sample that dominate the stacked signal, such as the two protoclusters from \citet{shi2021} that are expected to reach virial masses of the order of $10^{15}\,M_\odot$ at $z=0$. 

\section{Conclusion}

We analyse the convergence map reconstructed from ACT data at the location of \Planck\ protocluster candidates. To obtain a high enough signal to noise ratio, we stack the 1012 detections falling into ACT's footprint. We detect a signal corresponding to a peak in convergence with a mean value of 0.082, statistically different from stacks of 1012 random fields in ACT's footprint at a $12\,\sigma$ significance level. Due to the redshift sensitivity of CMB lensing, this can be safely interpreted as the \Planck\ protocluster candidates being correlated with the mass distribution at $z>1$, which is in line with previous photometric and spectroscopic redshift estimations for these objects. Taking into account the noise and a redshift range $1.5<z<3$, the integrated convergence over the whole \Planck\ beam yields a total mass of $(6.6\pm 1.4) \times 10^{14} M_\odot$. This value is in line with precedent estimations based on observations and simulations of subsamples of the catalogue from \citet{planck16}. It is also highly unlikely that such masses are virialised in the considered redshift range.

These points show that the \Planck\ catalogue of high redshift star-forming fields is correlated with the mass distribution at high redshift. This, together with their already known high star-formation rate, constitutes a solid argument in favour of this catalogue containing a notable amount of galaxy protoclusters.

\begin{acknowledgements}
TD acknowledges support from the France 2030 programme ANR-11-IDEX-0003, attributed through the
Astrophysics Axis of the Graduate School of Physics of Universit{\'e} Paris-Saclay. HD thanks Dr G. Hurier for an early preliminary unpublished study in 2020.
This paper made use of the ACT data publicly available at \url{https://lambda.gsfc.nasa.gov/product/act/actadv_prod_table.html}.
This paper made use of the \Planck\ data publicly available at \url{https://www.esa.int/Planck}. \Planck\ is a project of the European Space
Agency (ESA) with instruments provided by two scientific consortia funded by ESA member states and led by Principal Investigators from France and Italy, telescope reflectors provided through a collaboration between ESA and a scientific consortium led and funded by Denmark, and additional contributions from NASA (USA).
\end{acknowledgements}

\bibliography{cmb_lensing}

\appendix

\section{\label{appendix:cmb_lensing_density}Critical density for CMB lensing}

We display here the critical density for CMB lensing, for an observer at $z_{\rm obs}=0$ and a source at $z_s = z_{\rm CMB} = 1090$. The inverse of this function acts is the proportionality factor between the mass surface density and the convergence for a thin lens. We see that the variation in $\Sigma_c(z_l,z_{\rm CMB})$ is modest for $1.5<z_l<3$.

\begin{figure}[htbp!]
    \centering
    \includegraphics[width=0.95\hsize]{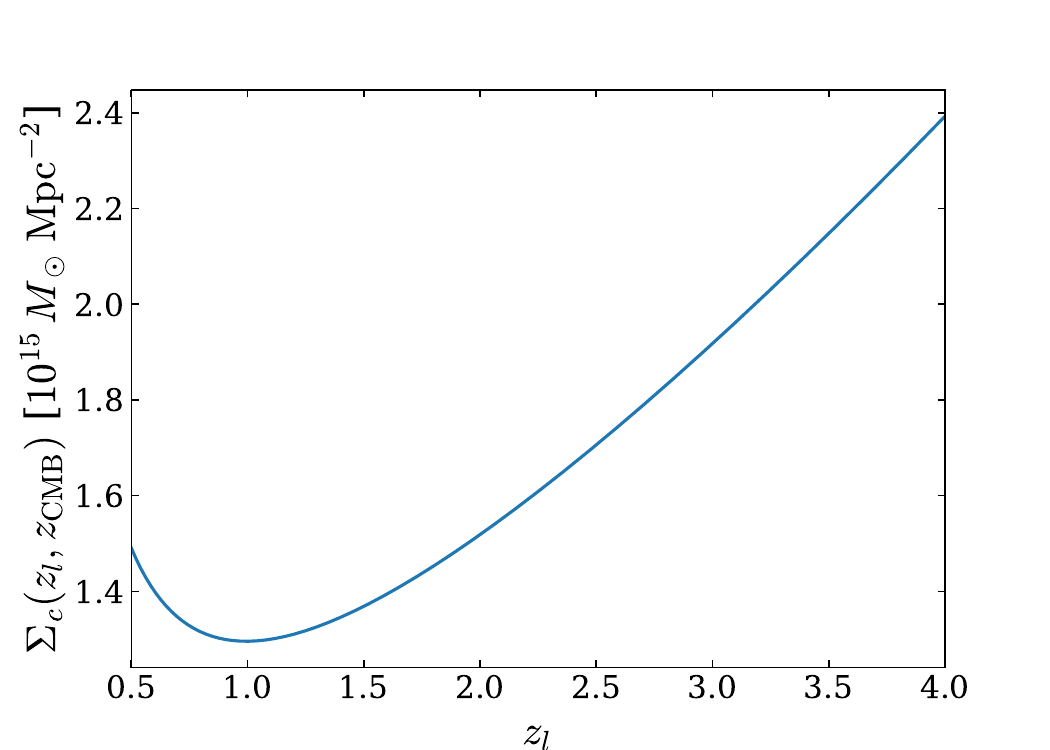}
    \caption{Evolution of the CMB lensing critical surface density $\Sigma_c(z_l,z_{\rm CMB})$ with respect to the redshift of the lens $z_l$ within the cosmological framework of \cite{planck2020_I,planck2020_VI}. The inverse of $\Sigma_c(z_l,z_{\rm CMB})$ acts as a weight for the detection of massive objects by CMB lensing, making it more sensitive to the mass distribution at high redshifts. The observer is at $z_{\rm obs}=0$ and the background source at $z_s = z_{\rm CMB} = 1090$.}
    \label{fig:crit_density}
\end{figure}

\section{\label{appendix:noise}Simulating mock noise maps}

\begin{figure}[htbp!]
	\centering
	\includegraphics[width=0.9\hsize]{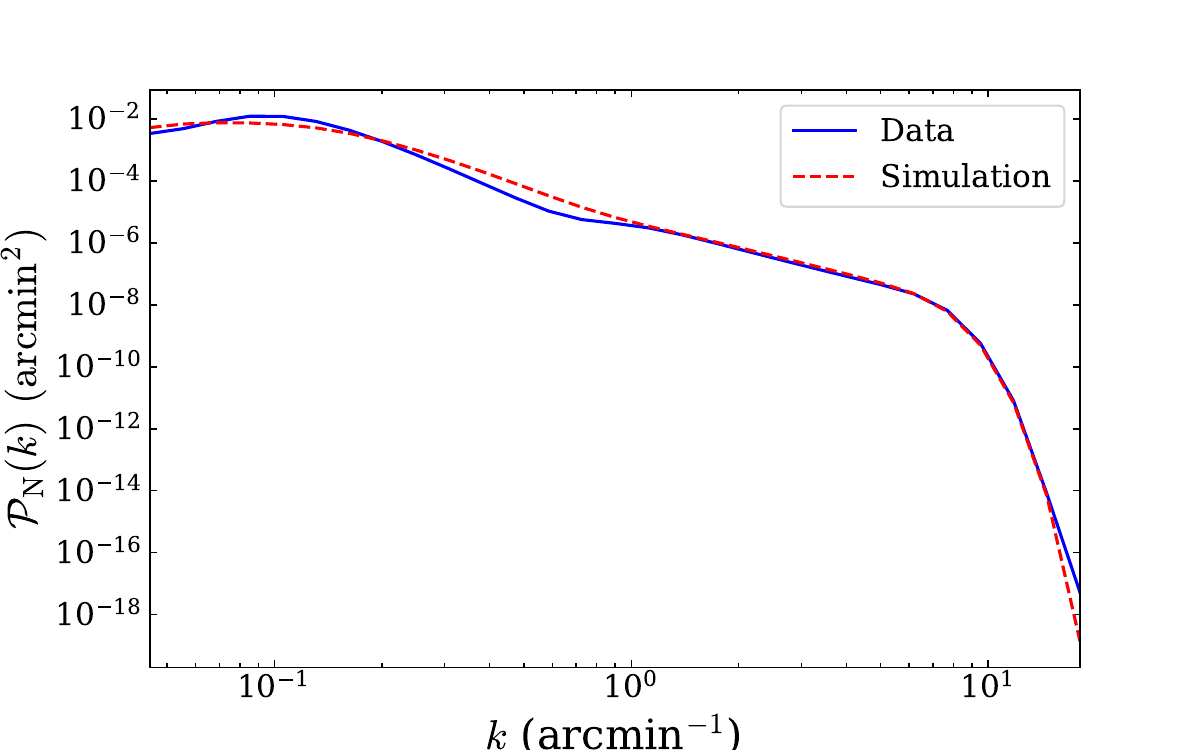}
	\caption{
		The power spectrum of the noise, taken as the zone outside of the centred disk of angular diameter $\ang{;10;}$ of the stacked convergence image, is plotted in blue. For comparison, we also display in red the power spectrum of a simulated noise map.
	}
	\label{fig:noise_PS}
\end{figure}

We show here how one can obtain an arbitrary number of random maps with the same statistics as the background noise. This noise is gaussian by hypothesis and its statistic is therefore entirely defined by either its variance or power spectrum $\mathcal{P}_{\rm N}$. We measure the latter employing the filtering method from \citet{arevalo2012}.

For a field $f$, we note $\FT{f}$ its Fourier transform, $\langle f \rangle$ its mean value, and $f^*$ its complex conjugate. Let $X_i$ be a realisation of a random white noise statistic, that is to say a gaussian noise without correlation (eq.~\ref{eq:white_noise})

\begin{equation}
	\langle X_i(\vect{x})^* X_i(\vect{x}+\vect{y}) \rangle_{\vect{x}} = \delta^{(2)} (\vect{y}) \, \cdot
	\label{eq:white_noise}
\end{equation}

We then define the field $\mathcal{S}_i$ according to eq.~\ref{eq:noise_simus} : 

\begin{equation}
	\mathcal{S}_i = \IFT{\FT{X_i} \times  \sqrt{\mathcal{P}_{\rm N}}} \, \cdot
	\label{eq:noise_simus}
\end{equation}

The power spectrum $\mathcal{P}_{S_i}$ of $\mathcal{S}_i$ is by definition the Fourier transform of its autocorrelation, or equivalently the squared amplitude of $\FT{S_i}$. As a real field with positive values, its square root is thus properly defined. By developing this relation, we get 

\begin{eqnarray}
	\label{eq:simus_ps_1}
	\mathcal{P}_{S_i} &=& \left| \FT{S_i}\right|^2 \\
	&=& \left| \FT{X_i}\sqrt{\mathcal{P}_{\rm N}} \right|^2 \\
	&=& \left| \FT{X_i}\right|^2 \mathcal{P}_{\rm N} \, \cdot
	\label{eq:simus_ps_2}
\end{eqnarray}

As $X_i$ is a white noise, $\left| \FT{X_i} \right|^2=1$ according to Eq.~\ref{eq:white_noise}. Therefore, $\mathcal{P}_{S_i} = \mathcal{P}_{\rm N}$, and the simulated maps have the same statistics as the background noise (Fig.~\ref{fig:noise_PS}).

\section{\label{appendix:other_maps}Mass estimations from other ACT lensing maps}

\begin{figure}[htbp!]
    \centering
    \includegraphics[width=0.9\hsize]{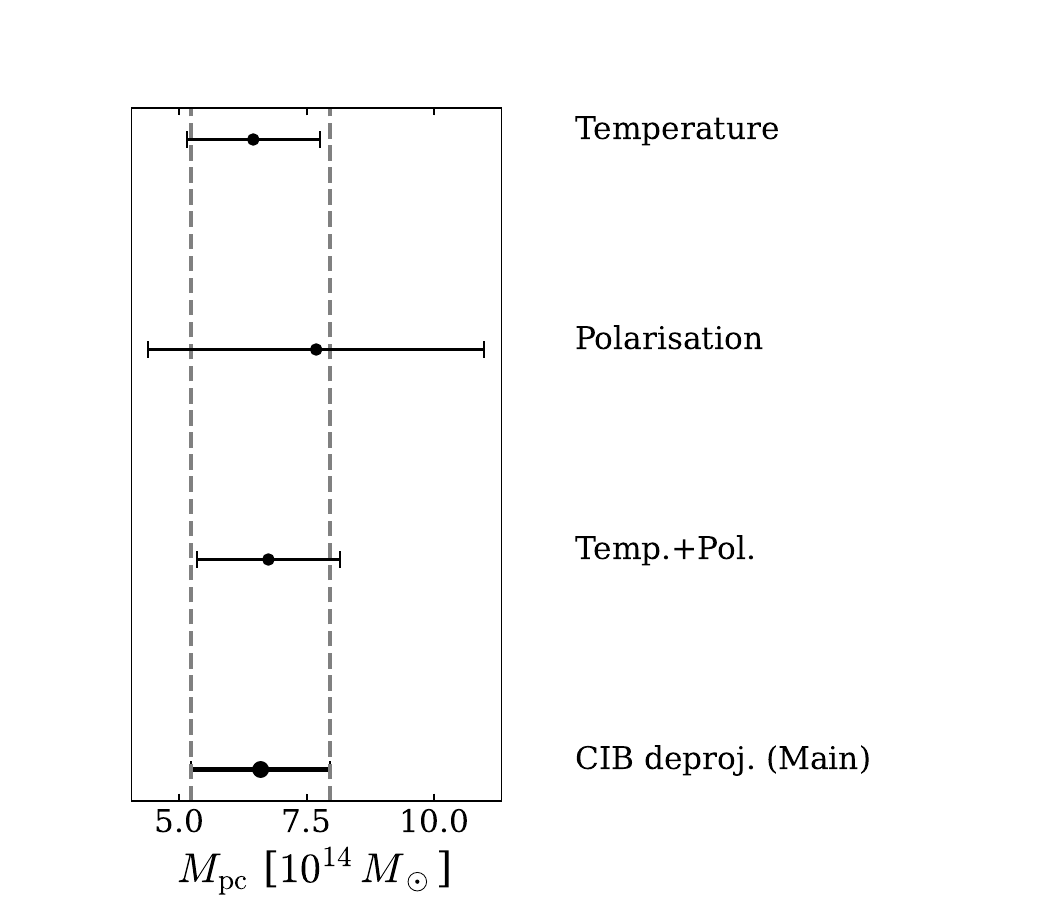}
    \caption{Masses estimated with the lensing maps reconstructed from temperature only, polarisation only, temperature and polarisation, and with a deprojection of the CIB (the one described in this work). The $1\,\sigma$ errorbars are represented as thin black lines. All estimates are consistent with each other.}
    \label{fig:different_masses}
\end{figure}

We specificed in Sect.~\ref{sect:methods} that we used the lensing map obtained from the CIB deprojected CMB maps of ACT. This is a natural choice, as \Planck\ protocluster candidates were initially discovered as specific spots of the CIB. However, one could ask if the systematics of the methods used to deproject the CIB might have a significant impact on our results. To ensure that we are not biased by this choice, we ran the same analysis as the one described in Sect.~\ref{sect:results_mass} on the lensing maps obtained from temperature only, polarisation only, and temperature and polarisation together. All these estimates are consistent with each other and with the estimation from CIB deprojected maps used in this work (Fig.~\ref{fig:different_masses}).

\section{\label{appendix:press_schechter}Press-Schechter formalism}

Here we recall the basics of the \citet{press1974}  formalism we used in our discussion in Sect.~\ref{sect:discussion_protocluster}. We place ourselves in the linear regime of cosmological perturbations and consider the matter overdensity field $\delta(\vect{x},z) = [\rho_{\rm m}(\vect{x},z)-\mean{\rho_{\rm m}}(z)]/ \mean{\rho_{\rm m}}(z)$, where $\rho_{\rm m}(\vect{x},z)$ is the matter mass density at redshift $z$ and position \vect{x} and $\mean{\rho_{\rm m}}(z)= 3 \Omega_{{\rm m},0}H_0^2(1+z)^3/8\pi G$ its mean value. The smoothing scale $R(M)$ is defined as the radius of a sphere of mass $M$ and constant density $\mean{\rho_{\rm m}}(z)$ (Eq.~\ref{eq:R_smooth})

\begin{equation}
    \label{eq:R_smooth}
    R(M,z) = \left[ \frac{3M}{4\pi \mean{\rho_{\rm m}}(z)}\right]^{1/3} \, \cdot
\end{equation}
 
We define the spherical top hat window function for mass $M$ (Eq.~\ref{eq:window}), which in turn allows us to properly define the smoothed overdensity field at the scale corresponding to $M$ (Eq.~\ref{eq:smoothed_overdensity})

\begin{equation}
    \label{eq:window}
    \mathcal{W}(\vect{x},M,z) = \left\{
    \begin{array}{ccc}
       \dfrac{3}{4\pi R(M,z)^3}  & {\rm if} & \vert\vect{x}\vert \leq R(M,z)\\
        0 & {\rm if} & \vert\vect{x}\vert > R(M,z)
    \end{array}
    \right.
\end{equation}

\begin{equation}
    \label{eq:smoothed_overdensity}
    \delta_M(\vect{x},z) = \int \mathcal{W}(\vect{x}-\vect{y},M,z) \delta(\vect{y},z) {\rm d}\vect{y}^3 \, \cdot
\end{equation}

In the Press-Schechter formalism, $\delta_M$ is a Gaussian random field the variance of which can be expressed with respect to the spatial Fourier transform $\FT{\mathcal{W}}(k,M,z)$ of $\mathcal{W}(\vect{x},M,z)$, that is isotropic in the specific case of the top-hat filter, and the matter power spectrum $\mathcal{P}_{\rm m}$ (Eq.~\ref{eq:matter_variance}).

\begin{equation}
    \label{eq:matter_variance}
    \sigma^2(M,z) =  \frac{1}{2\pi^2}\int_0^{+\infty} k^2 \vert \FT{\mathcal{W}}(k,M,z)\vert^2 \mathcal{P}_{\rm m}(k,z) {\rm d} k \, \cdot
\end{equation}

As $\delta_M$ is a gaussian random field, the probability distribution for an overdensity $\delta_M$ of mass at least $M$, knowing that $\delta_M>0$, is

\begin{equation}
    \label{eq:proba_deltam}
    \varrho(\delta_M,M,z) = \sqrt{\frac{2}{\pi}} \frac{1}{\sigma(M,z)}\exp\left[ -\frac{\delta_M^2}{2 \sigma^2(M,z)} \right] \, \cdot
\end{equation}

We consider that a halo forms if the linear overdensity has crossed the critical collapsing threshold $\delta_c \simeq 1.686$ in an Einstein-de Sitter Universe, which is an excellent approximation of our own Universe at redshifts of interest for this work. Therefore, the probability that a halo of mass superior or equal to $M$ has formed at redshift $z$ is 

\begin{equation}
    \label{eq:proba_halo}
    p(M,z) = \int_{\delta_c}^{+\infty} \varrho(\delta_M,M,z) {\rm d} \delta_M = {\rm erfc} \left[ \frac{\delta_c}{\sqrt{2}\sigma(M,z)} \right]
\end{equation}
with ${\rm erfc}$ the complementary error function.

\end{document}